\documentclass{iaus}

\usepackage[latin2]{inputenc}
\usepackage{amsmath}
\usepackage{ccfonts}
\usepackage[english]{babel}
\usepackage{graphicx}

\newcommand{\bm}[1]{\mbox{{\boldmath $#1$}}}

\newcommand{\pder}[3]{\frac{{\partial}^{#3} {#1}}{{\partial} {#2}^{#3}}}
\newcommand{\pc}{\textrm{ pc}}
\newcommand{\kpc}{\textrm{ kpc}}

\newcommand{\Myr}{\textrm{ Myr}}
\newcommand{\Gyr}{\textrm{ Gyr}}
\newcommand{\cm}{\textrm{ cm}}

\newcommand{\km}{\textrm{ km}}
\newcommand{\s}{\textrm{ s}}
\newcommand{\erg}{\textrm{ erg}}

\newcommand{\muG}{\mu{\textrm{G}}}

\newcommand{\G}{\textrm{ G}}

\title[Cosmic-ray driven dynamo]{Cosmic-ray driven dynamo \\
       in galactic disks}

\author[Hanasz et al.]
             {M.  Hanasz$^1$, 
	      K. Otmianowska-Mazur$^2$,
	      H. Lesch$^3$,
	      G. Kowal$^{2,4}$,\\ 
	      M. Soida$^2$,  
	      D. Wóltañski$^1$ 
	      K. Kowalik$^1$,  
	      R.K. Paw³aszek$^1$ \\ 
	      B. Kulesza-\.Zydzik$^2$}

\affiliation{$^1$Centre for Astronomy, Nicholas Copernicus University,
  PL-87148 Piwnice/Toru\'n, Poland, mhanasz@astri.uni.torun.pl
           $^2$Astronomical Observatory, Jagiellonian University,
           ul. Orla 171, 30-244 Krak\'ow,
           $^3$Astronomical Observatory, Munich University,
           Scheinerstr. 1, D-81679, Germany,
           $^4$Department of Physics and Astronomy, McMaster University,
           1280 Main St. W., Hamilton, ON L8S 4M1, Canada.
	   }

\pubyear{2009}
\volume{259}  
\pagerange{100--100}
\date{"YOUR MAILING DATE"  and in revised form ??}
 \jname{Cosmic Magnetic Fields: From Planets,
to Stars and Galaxies} \editors{K.G. Strassmeier, A.G. Kosovichev \&
J.E. Beckman, eds.}

\begin{document}
\maketitle
\begin{abstract}
We present new developments on the Cosmic--Ray driven, galactic dynamo, modeled
by means of direct, resistive CR--MHD simulations, performed with ZEUS and
PIERNIK codes. The dynamo action, leading to the amplification of large--scale
galactic magnetic fields on galactic rotation timescales, appears as a result of
galactic differential rotation, buoyancy of the cosmic ray component and
resistive dissipation of small--scale turbulent magnetic fields. Our new results
include demonstration of the global--galactic dynamo action driven by Cosmic Rays
supplied in supernova remnants. An essential outcome of the new series of global
galactic dynamo models is the equipartition of the gas turbulent energy with
magnetic field energy and cosmic ray energy, in saturated states of the dynamo
on large galactic scales.
\end{abstract}
%
%
%
%
%
\section{Introduction}
Numerous processes have been proposed to explain initial magnetic fields in
early galaxies. However all these processes, like phase transitions in the early
universe or Biermann battery in protogalactic objects provide only very week
magnetic fields of the order of $ B \sim 10^{-20}  \G$ at the beginning of
galactic evolution. On the other hand, Rees (\cite{rees-87}) proposed that
initial magnetic fields in galaxies might have been generated in first stars and
then scattered in the interstellar medium (ISM) by SN explosions, and
subsequently amplified in plerionic (Crab--type) supernova remnants (SNRs). The
mean magnetic field on the galactic scale has been estimated to be of the order
of  $ 10^{-9} \G$. Radio observations (see Beck \cite{beck-09}, this volume)
indicate that typical contemporary magnetic fields in spiral galaxies are of the
order of few up to few tens of $\muG$, which means that magnetic fields have
undergone amplification by at least four orders of magnitudes within the
galactic lifetime. Therefore, a model for efficient magnetic field amplification
during galactic evolution is necessary. The standard model of magnetic field
amplification in disk galaxies is based on the  theory of turbulent mean field
dynamo  (see Widrow \cite{widrow-02} for a recent review). 
%
%
%
%
\par The mean--field dynamo theory has been successful in explaining magnetic
fields in various astrophysical objects, however, the classical, kinematic
dynamo seems to be rather slow in galaxies.  Magnetic field amplification
timescale  $t_{\rm dynamo} \sim (0.5 \div 1) \times 10^9\textrm{ yrs}$ is too long to
explain $\sim 1 \muG$ magnetic fields in galaxies which are only few $10^9$ yrs
old ($z \sim 1\div3$)  (Wolfe, Lanzetta \& Oren \cite{wolfe-etal-92}, Kronberg
et al. \cite{kronberg-etal-08}). These circumstances seem to indicate a need of
an alternative approach.  We suggest that direct numerical MHD modeling  of ISM
dynamics, through considerations of detailed physical processes, involving major ISM
components: gas, magnetic fields and cosmic rays, is necessary to follow magnetic
history of galaxies. 
\section{Cosmic Rays in the interstellar medium}
%
%
The dynamical role of CRs was recognized for the first time by  Parker
(\cite{parker-66}), who noticed that a vertically stratified ISM, consisting of
thermal gas magnetic field and CRs is unstable due to buoyancy of the weightless
components: magnetic fields and CRs.  
The CR component appears to be an important ingredient of this process. According to the
diffusive shock acceleration models CRs are continuously supplied to ISM by SN
remnants. Therefore, strong buoyancy effects due to CRs are unavoidable.
\par Theories of diffusive shock acceleration  predict that about 10 \% of the
$\sim 10^{51} \erg$ of the SN II explosion energy is converted to the CR energy
(see e.g. Jones et al \cite{jones-etal-98} and references therein). 
Observational data indicate that gas, magnetic fields and CRs  appear in
approximate energetic equipartition, which means that all the three components
are dynamically coupled. Moreover, numerical experiments by Giaccalone \&
Jokipii (\cite{giaccalone-jokipii-99})  indicate that CRs diffuse
anisotropically in the ISM, along the mean magnetic field direction. 
%
%
%
%
\section{Cosmic Ray transport}
To incorporate CR propagation in MHD considerations we use the diffusion -
advection equation (e.g. Schlickeiser \& Lerche \cite{schlickeiser-lerche-85})
\begin{eqnarray}
\pder{e_{\rm cr}}{t}{}+ \bm{\nabla} (e_{\rm cr} \bm{v}) 
                  = -p_{\rm cr} \bm{\nabla}\cdot \bm{v}
             +{\bm{\nabla} (\hat{K} \bm{\nabla} e_{\rm cr})}
	     \label{eqn:crtransport}\\
      \mbox{{ + CR \ sources (SN remnants)}} \nonumber
\end{eqnarray}
where the diffusion term in written in the tensorial form 
to account for anisotropic diffusivity of CRs. The sources of CRs on the
rhs. of  Eqn.~\ref{eqn:crtransport} correspond to the CR production in Supernova
remnants. 
\par The dynamics of ISM including the energetics of CRs can be now described by the
CR transport Eqn.~(\ref{eqn:crtransport}) and the system of MHD equations. We
note that in the presence of cosmic rays additional source term:  $-\nabla
P_{\rm CR}$ should be included in the gas equation of motion (see e.g. Berezinski
et al. \cite{berezinski-etal-90}), in order to incorporate the effects of CRs on
gas dynamics. The cosmic ray diffusion--advection Eqn.~(\ref{eqn:crtransport})
has been supplemented to the MHD algorithm  of Zeus--3D MHD code by Hanasz \&
Lesch (\cite{hanasz-lesch-03a}) with the aim of studying ISM physical processes,
 e.g. Parker instability, in which CRs play a dynamically important role.  

\section{CR-driven dynamo}
Referring back to the issue of galactic magnetic fields, we are primarily
interested in processes leading to the amplification of large scale magnetic
fields. The original idea of CR--driven dynamo has been raised by  Parker
(\cite{parker-92}). Our first local (shearing--box) CR--driven dynamo numerical
experiments (Hanasz et al. \cite{hanasz-etal-04}, \cite{hanasz-etal-06};
Otmianowska-Mazur et al. \cite{otmianowska-mazur-etal-07}) rely on
the following ingredients: (1.)  the cosmic ray component described by the 
diffusion--advection transport equation ~(\ref{eqn:crtransport}),  including 
localized sources of cosmic rays ---  supernovae remnants, exploding randomly in
the disk  volume,  while SN shocks and thermal effects are neglected  (see
Gressel et al. \cite{gressel-etal-08a}, \cite{gressel-etal-08b}) for a recent dynamo model relying on the thermal energy
output from SNe);   (2.) resistivity of the ISM (see Hanasz, Otmianowska--Mazur
and Lesch \cite{hanasz-etal-02}), leading to  topological evolution of magnetic
fields;    (3) shearing boundary conditions (Hawley, Gammie and  Balbus
\cite{hawley-etal-95}) together with Coriolis and tidal forces, aimed at
modeling of differentially  rotating  disks in the local approximation;  (4)
realistic vertical disk gravity and rotation following the model of ISM in the
Milky Way by Ferri\`ere (\cite{ferriere-98}).
\par
\begin{figure}[!h]
\centerline{\includegraphics[width=0.4\textwidth]{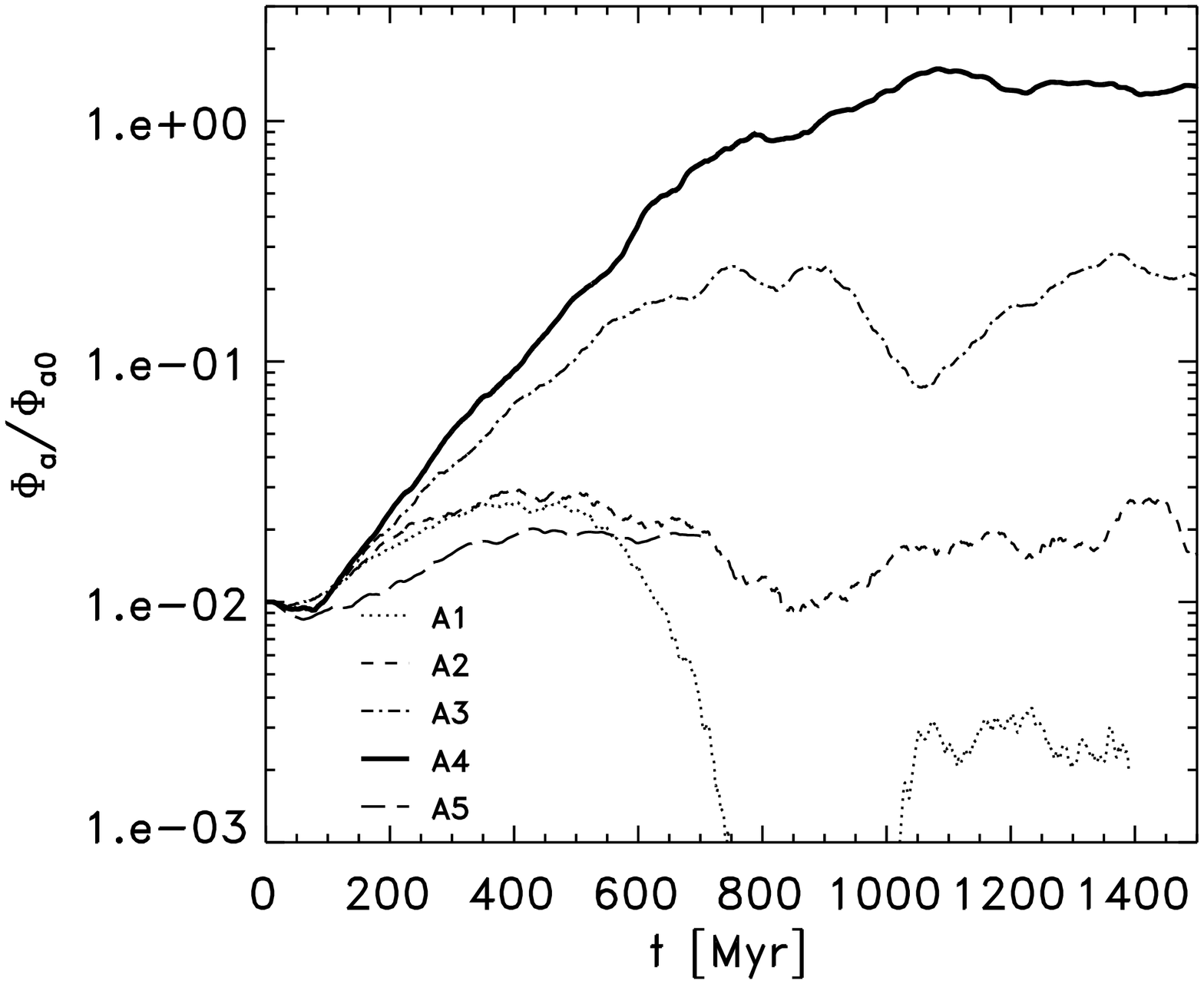}
\qquad \includegraphics[width=0.4\textwidth]{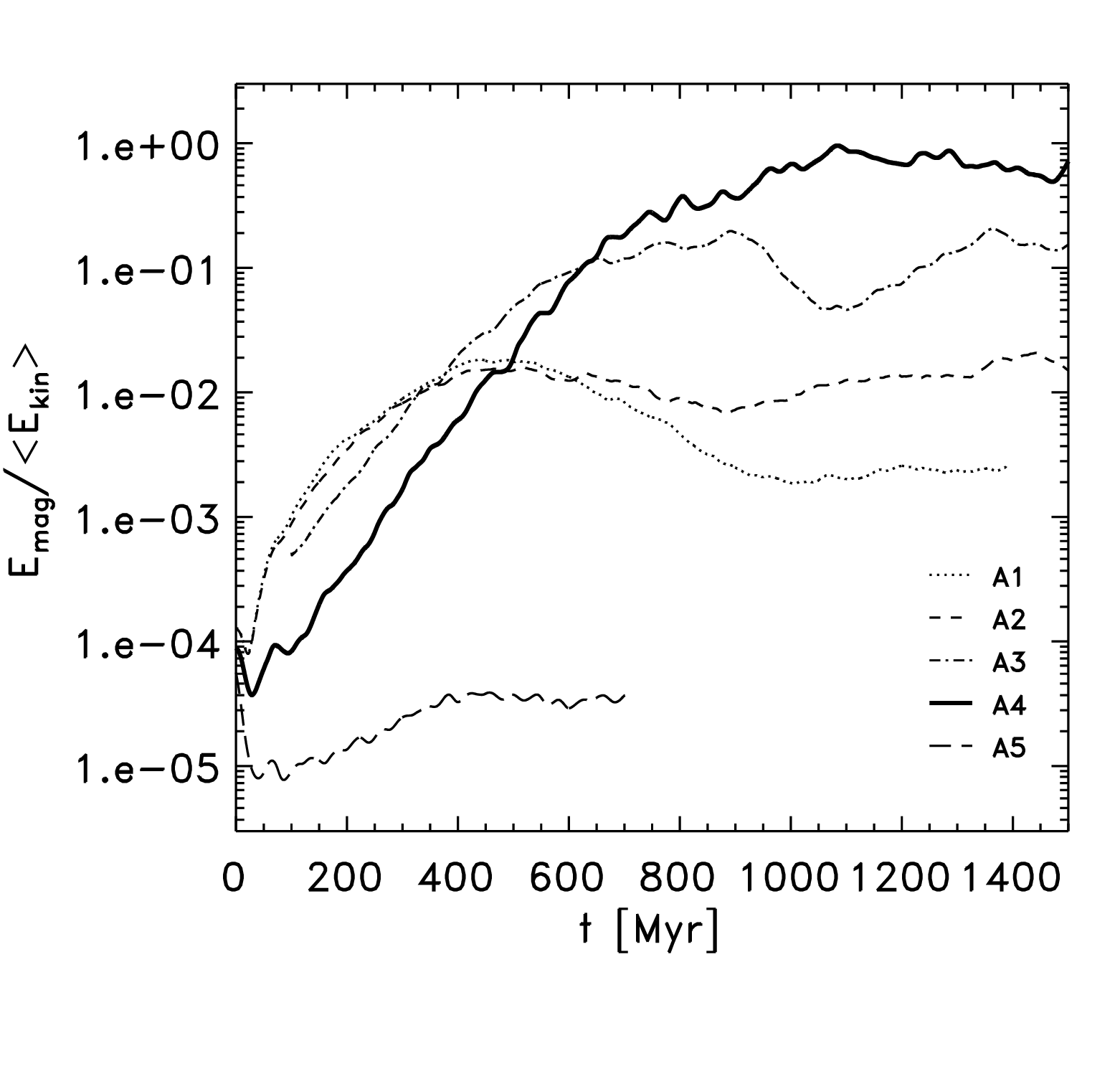} }
\caption{
Time evolution of azimuthal  magnetic flux and total magnetic energy
for different values of magnetic diffusivity in simulation series A. The curves represent
respectively cases of $\eta=0$ (A1), $\eta=1$ (A2), $\eta=10$ (A3), $\eta=100$ (A4) 
and $\eta=1000$ (A5) in units $\pc^2 \Myr^{-1}$.}
\label{fig:mag-diff}
\end{figure}

 In the most recent  paper (Hanasz et al. \cite{hanasz-etal-09}) we present a
parameter study of the CR--driven dynamo model by examining the dependence of
magnetic field amplification on magnetic diffusivity, supernovae rate determining
the CR injection rate, temporal modulation of SN activity, grid resolution, and
CR diffusion coefficients. We find the dominating influence of magnetic
diffusivity (treated as a free parameter), among other parameters, 
(Fig.~\ref{fig:mag-diff}) on the efficiency of magnetic field amplification. 
\par The fastest magnetic field amplification, observed in the experiments, 
coincides with the magnetic diffusivity comparable to $\eta  \simeq
\frac{1}{3}\eta_{turb}(obs)$, where $\eta_{turb}(obs) \simeq\frac{1}{3} 100 \pc
\times 10 \km\s^{-1}\simeq 10^{26} \cm \s^{-1}$. The question of which physical
process can be responsible for that large magnetic diffusivity coefficients
points out  towards the current studies on magnetic reconnection  (Lazarian et al.
\cite{lazarian-etal-04}, Cassak et al. \cite{cossak-etal-06})
\par
\begin{figure}[!h]
\centerline{\includegraphics[width=0.4\columnwidth]{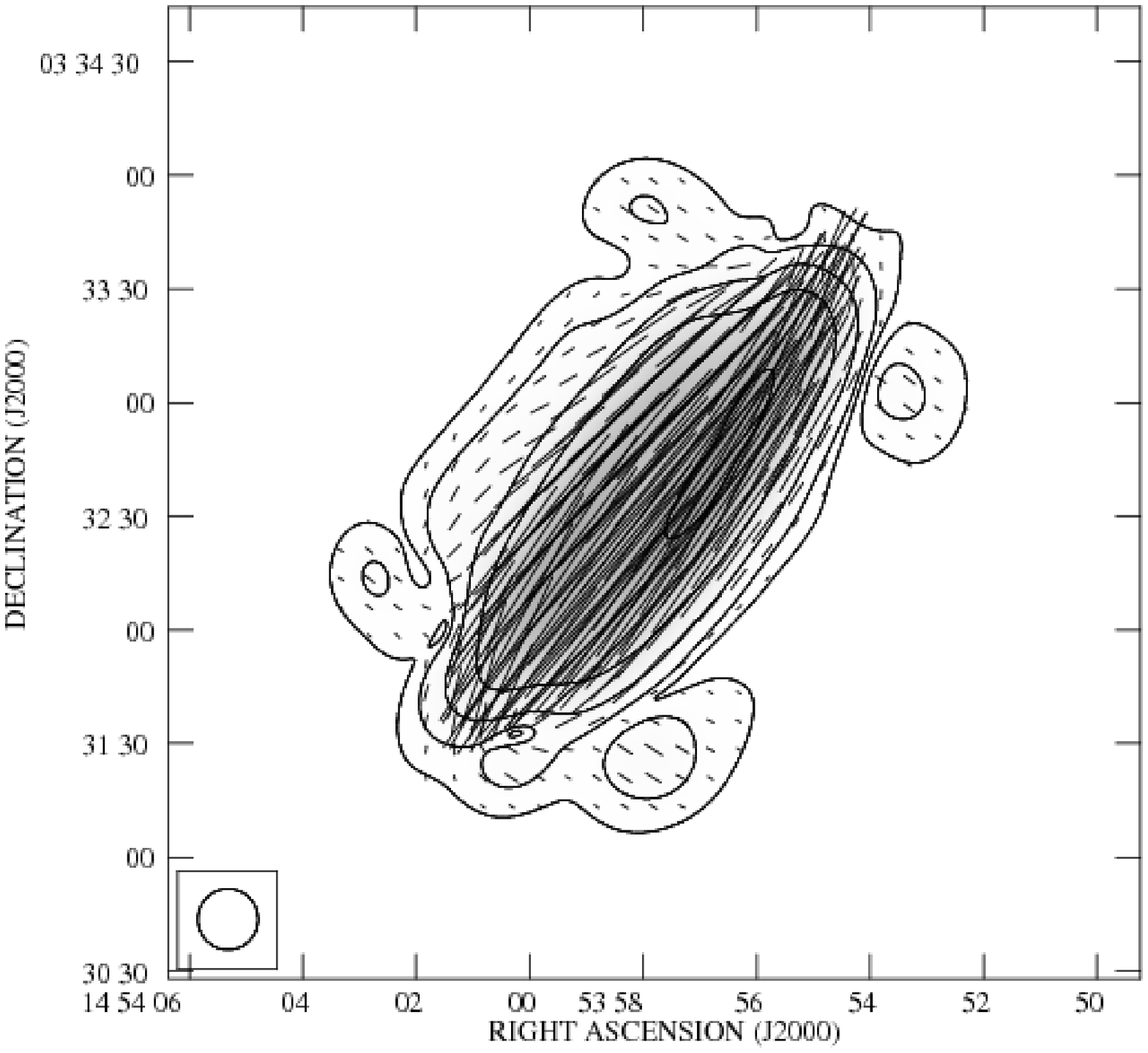}
\includegraphics[width=0.4\columnwidth]{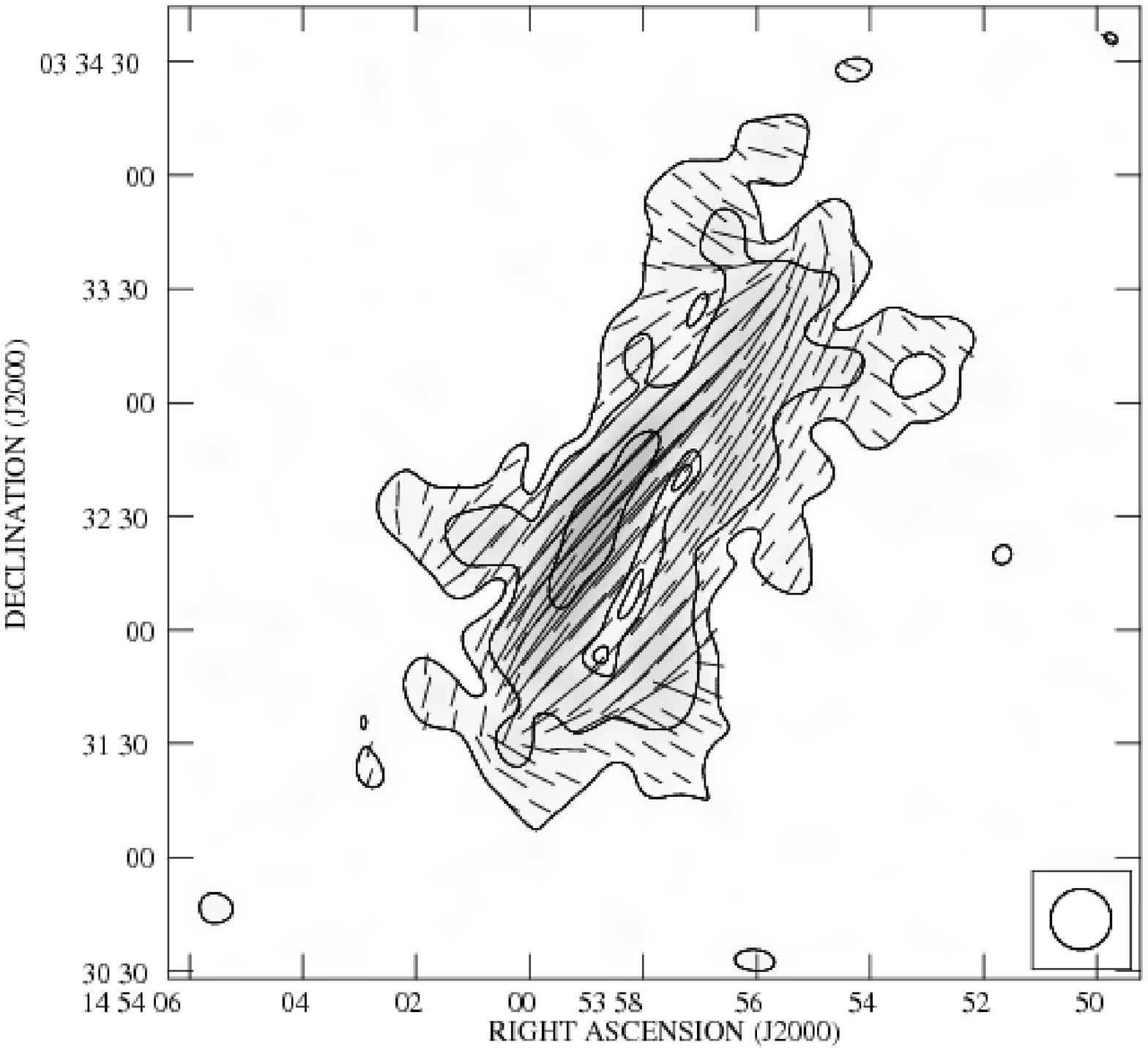}} 
\label{fig:radiomap}
\caption{Synthetic radio--map of polarized radio emission for our model at 
the inclination $i = 80^\circ$ and a map of real galaxy NGC 5775 } 
\end{figure}

 To investigate observational properties of current dynamo models we perform a
series of numerical simulations, placing the local shearing--boxes at different
galactocentric radii. We replicate the contents of local cubes  into rings, and
then combine rings to build up a disk (Otmianowska--Mazur et al.
\cite{otmianowska-mazur-etal-09}).  A synthetic polarization radio--map of
synchrotron emission constructed, on the base of simulated magnetic fields and
CR distribution is shown in Fig.~\ref{fig:radiomap}.  

A polarized
emission map together with superposed polarization vectors of synchrotron
emission of an edge--on galaxy NGC 5775 is shown for comparison. As it is
apparent the  $X$--shaped structures of magnetic field, present in
the real galaxy, form also in the synthetic radio image. The presented results
demonstrate that the CR--MHD simulations provide a way of verification of
CR--dynamo models against the observational data.
\par To summarize the overall outcome of the local simulations of CR--driven dynamo,
we note that the model provides efficient amplification of large--scale
magnetic fields, which attain, in the saturated state, equipartition with
turbulent gas motions. However,we note also a drawback of the local models,
since  the CR energy density remains more than an order of magnitude larger 
than gas and magnetic energy densities. We recognize that improvements of the
current models are necessary, since the possible reason of the excess of cosmic
rays may result from the effect of trapping of cosmic rays by predominantly
horizontal magnetic field in the computational box which is periodic in
horizontal directions. Therefore, global galactic disk simulations should be
performed to enable escape of CRs along the predominantly horizontal magnetic field 
in the galactic
plane, in order to avoid the CR excess.
%
%
%
\section{Global disk simulations}
Recently we started a new series of global simulations of CR--driven dynamo  with
the aid of  PIERNIK MHD code (see Hanasz et al. \cite{hanasz-etal-09a},
\cite{hanasz-etal-09b}, \cite{hanasz-etal-09c}, \cite{hanasz-etal-09d}),
which  is a grid--based MPI parallelized, resistive MHD code based on the
Relaxing TVD (RTVD) scheme by Jin \& Xin~(\cite{jin-95}) and Pen et
al.~(\cite{pen-03}).  The original scheme is extended to deal with the diffusive
CR component  (see Hanasz \& Lesch~\cite{hanasz-lesch-03a}).
\par
\begin{figure}
 \centerline{
  \includegraphics[width=0.19\columnwidth]{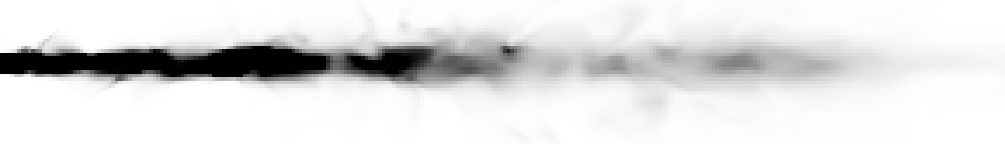}
  \includegraphics[width=0.19\columnwidth]{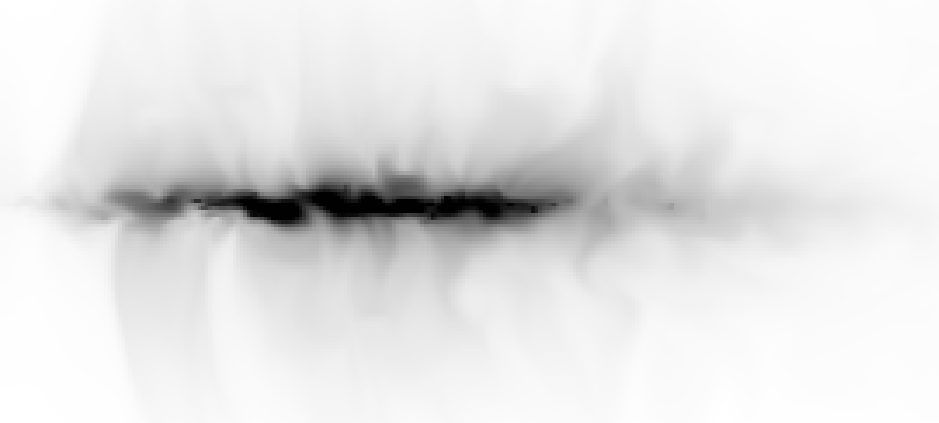}
  \includegraphics[width=0.19\columnwidth]{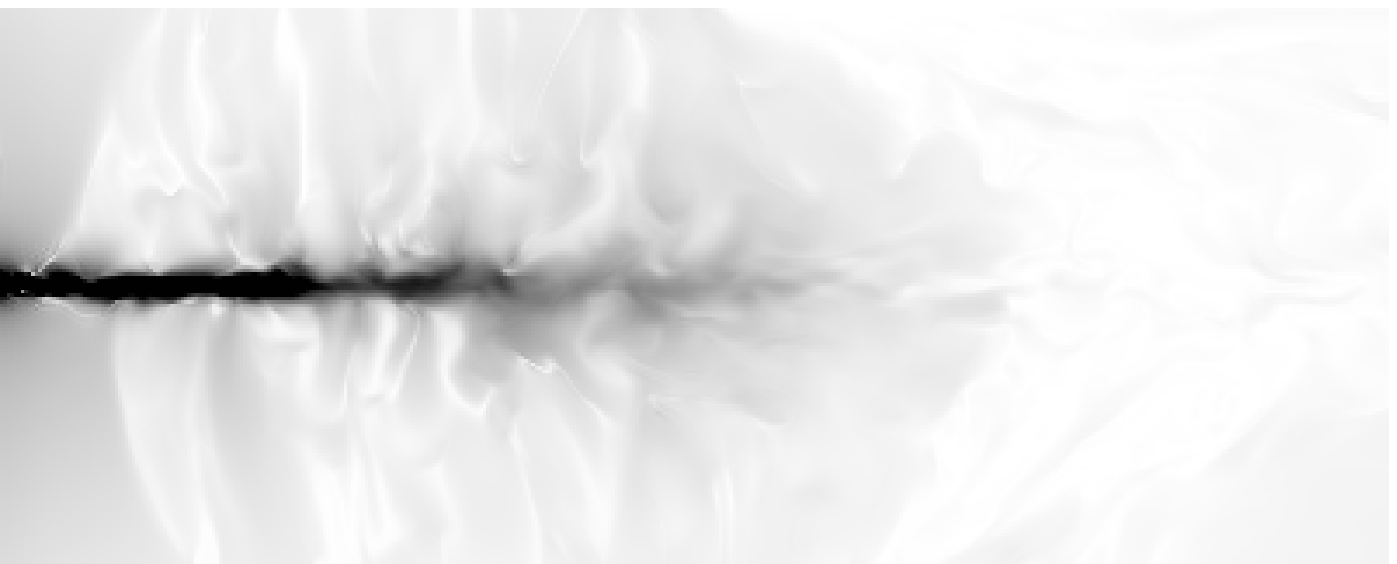}
  \includegraphics[width=0.19\columnwidth]{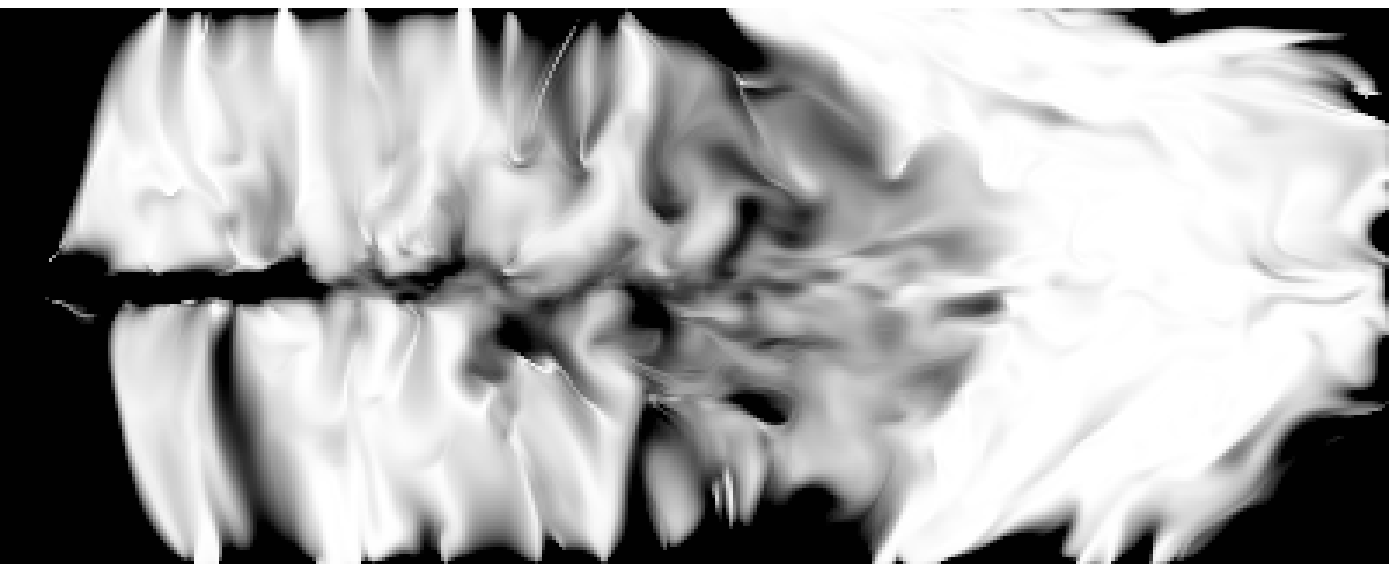}
  \includegraphics[width=0.19\columnwidth]{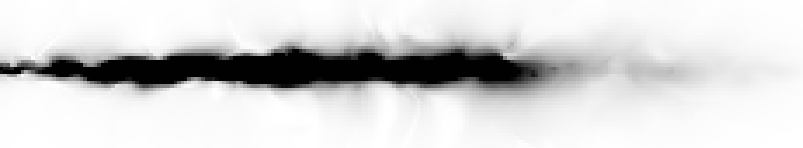}}
 \centerline{  
  \includegraphics[width=0.19\columnwidth]{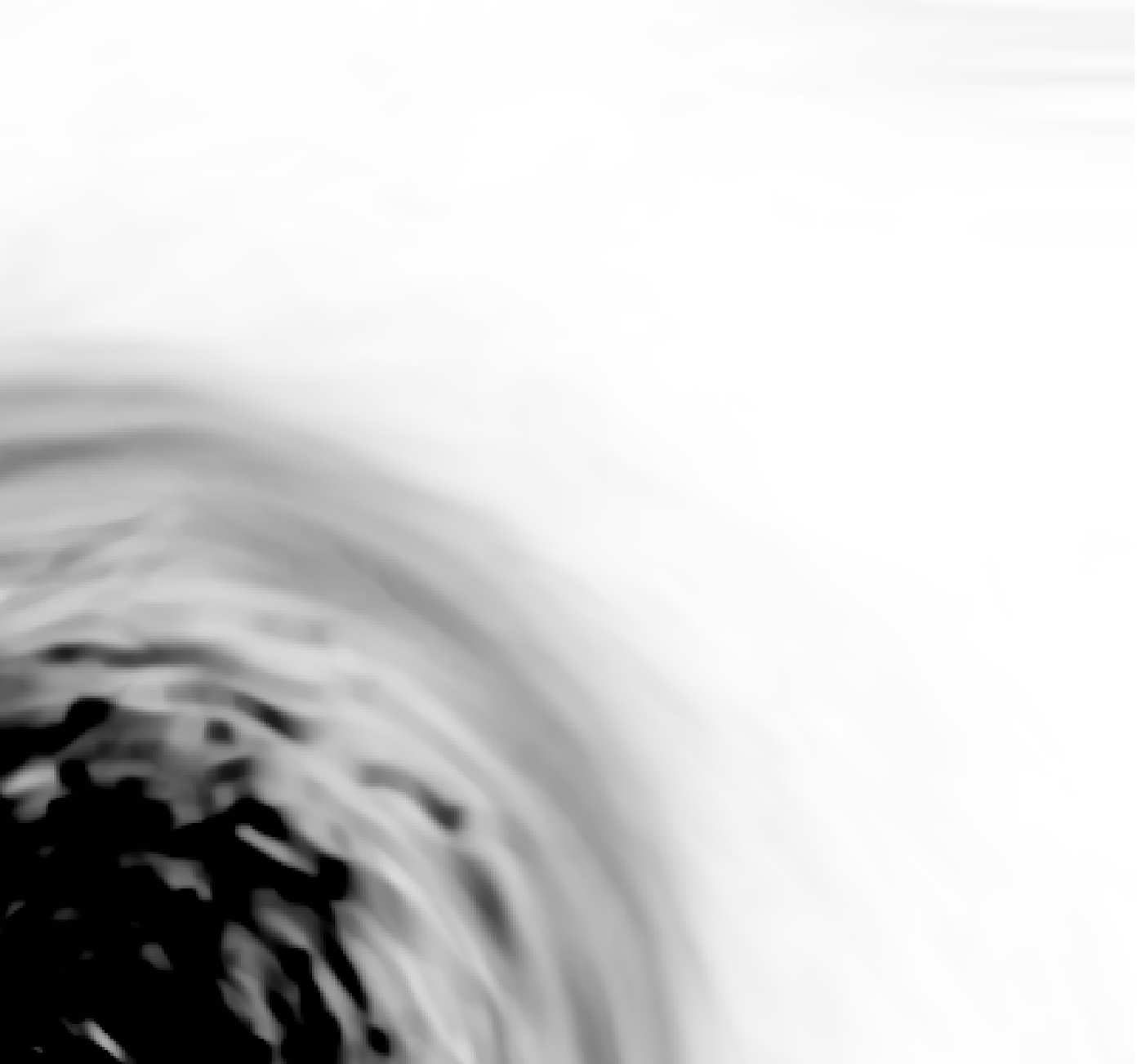}
  \includegraphics[width=0.19\columnwidth]{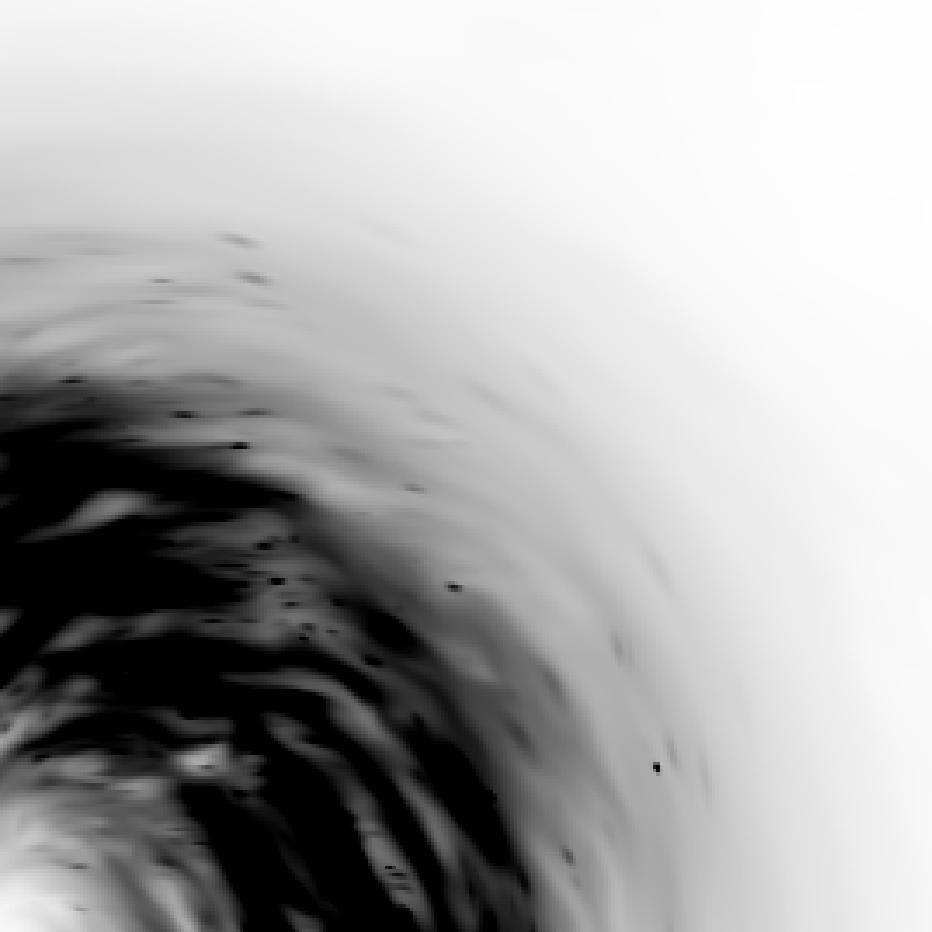}
  \includegraphics[width=0.19\columnwidth]{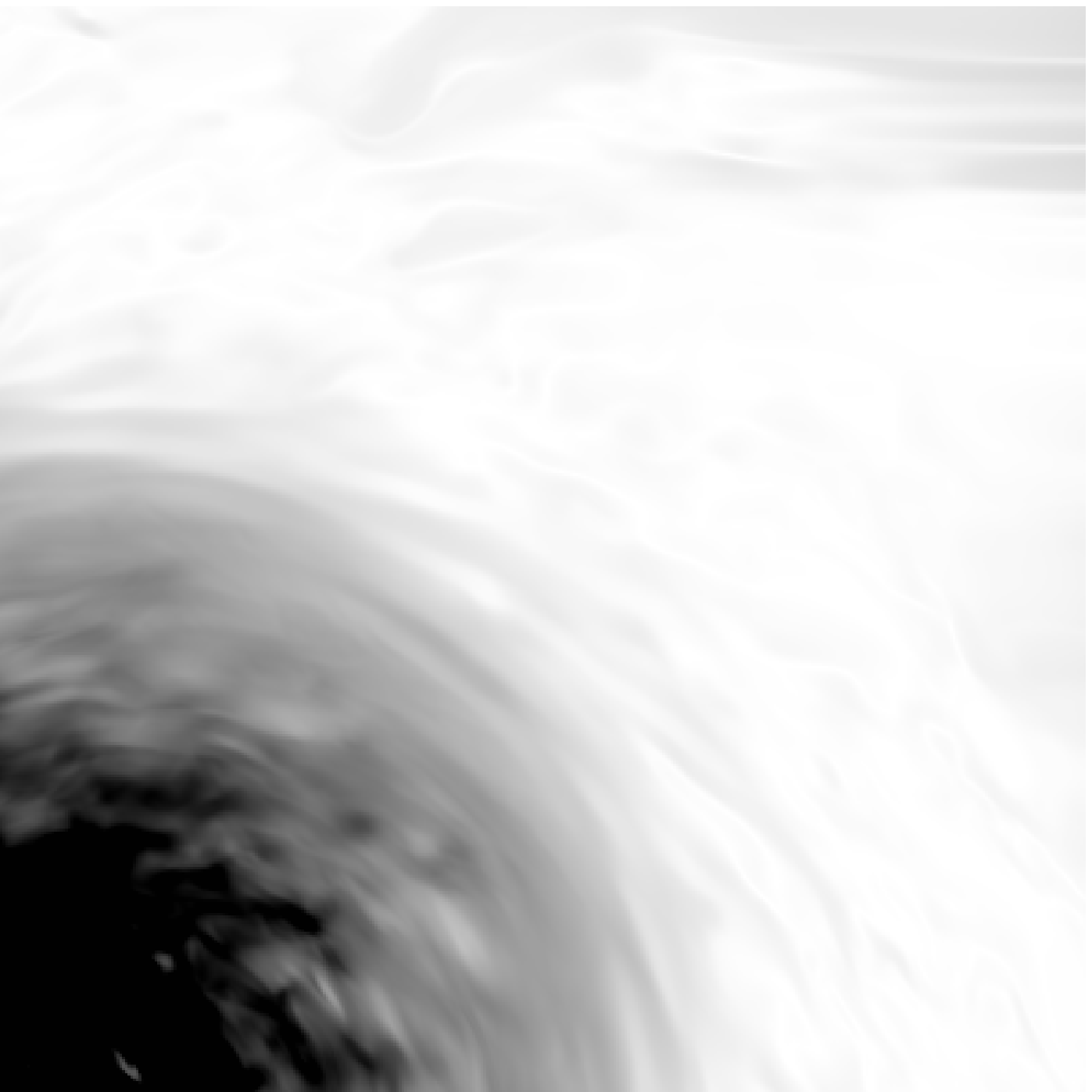}
  \includegraphics[width=0.19\columnwidth]{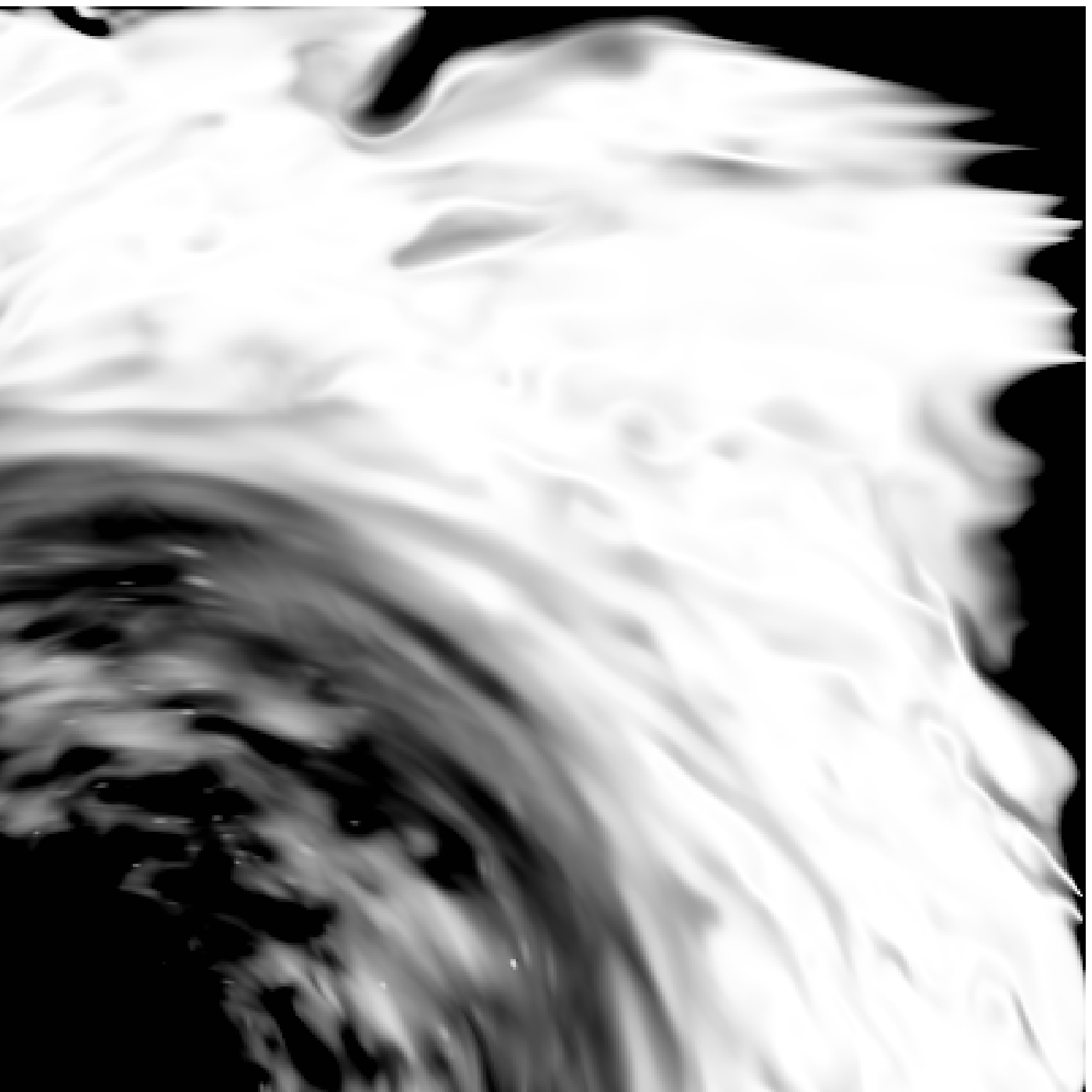}
  \includegraphics[width=0.19\columnwidth]{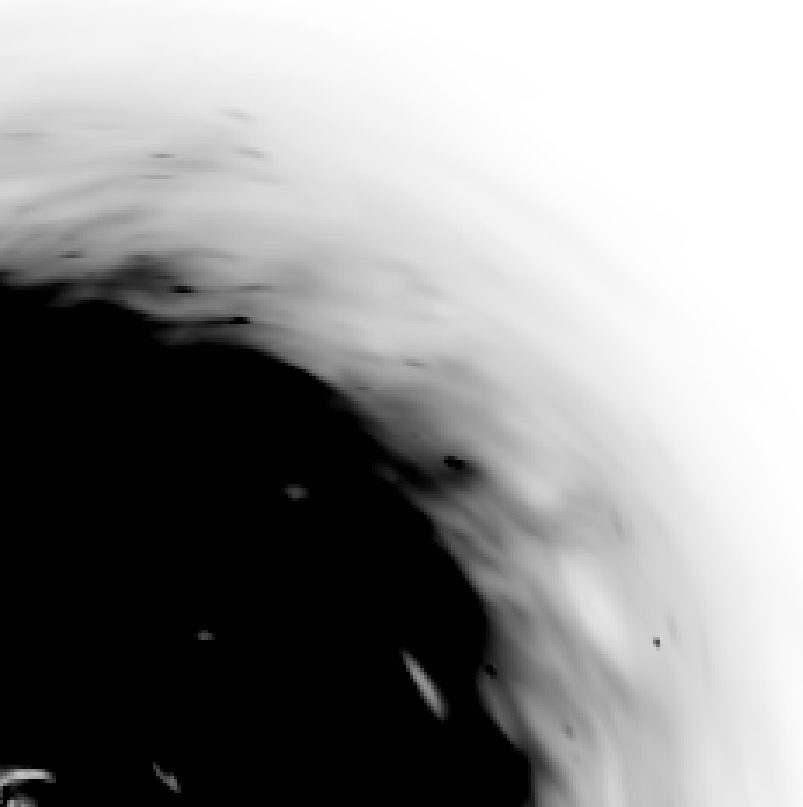}} 
\caption{Horizontal and vertical slices through the computational box for a
simulation of CR--driven dynamo in a semi--global galactic disk. The color maps
reflect: 1: gas density, 2: CR energy density, 3: magnetic field strength, 4.
the ratio of magnetic to CR energy densities and 5. synchrotron emissivity for a
saturated state of the dynamo. The gray--scales in all panels, except 4. are
chosen arbitrarily. The darker shades correspond to larger values of displayed 
quantities. The greyscale of panel 4. is such that black areas represent
values of $e_{\rm mag}/e_{\rm cr}\simeq 1$.}
\label{fig:globaldisk}
\end{figure}

 In Fig.~\ref{fig:globaldisk} we show results of one of the first semi--global
simulations of CR--driven dynamo. The simulations have been performed for a
quarter of galactic disk resembling Milky Way, with parallel CR diffusion
coefficient $K_\parallel = 3\times 10^{28} \cm^2\s^{-1}$,  in a computational
domain  $20 \kpc \times 20 \kpc \times 20 \kpc $,  and resolution of $400
\times 400 \times 160 $ grid cells, in 16 MPI blocks.

\par The semi--global simulations confirm the fast amplification rate of the mean
magnetic field: growth time of the mean magnetic flux is approximately 210 Myr
in the whole galaxy, which is close to the galactic rotation period near the
orbit of Sun. As expected, the global simulations introduce a significant
improvement with respect to the former shearing--box simulations. It is  apparent in
Fig.~\ref{fig:globaldisk} that the ratio of magnetic to cosmic ray energy is
close to one in the synchrotron emitting volume  (compare panels 4.and 5.),
while over--equipartition of CR energy with respect to magnetic energy still
holds  outside the disk.
\par The latest development of our CR--dynamo model is a fully global galactic disk
simulation (see the complementary paper by Hanasz, et al. this volume), where we demonstrate that dipolar magnetic fields
supplied on small SN--remnant scales, can be amplified exponentially by the
CR--driven dynamo to the present equipartition values, and transformed
simultaneously to large galactic--scales. 
\section{Conclusions}
We have shown that the CR contribution to the dynamics of ISM, studied by means of CR--MHD simulations,
in both local and global scales, leads to a very efficient magnetic field
amplification in galactic disks.
\par The cosmic ray driven dynamo amplifies efficiently galactic magnetic fields
on a ti\-me\-sca\-le of galactic rotation.  The resulting growth of the
large--scale magnetic field by 4 orders of magnitude within $~ 2 \Gyr$ (Hanasz
at al. \cite{hanasz-etal-04}), is fast enough to expect $\sim 1 \muG$ magnetic
field in galaxies at $z \sim 1\div 3$.
\par We point out the  advantage of global disk models, which on the contrary to
shearing--box models, provide CR  energy equipartition with magnetic field in the
synchrotron emitting part of the disks.
\subsection*{Acknowledgements}
This work was supported by Polish Ministry of
Science and Higher Education through the grants 92/N--ASTROSIM/2008/0 
and \mbox{PB 0656/P03D/2004/26} and
by Nicolaus Copernicus University through Rector's grant No. 516--A
%
%
%
%
%
%
%
%

\end{document}